\documentclass{achemso}
\usepackage[version=3]{mhchem} 
\usepackage[T1]{fontenc}
\usepackage{lmodern}
\usepackage{amsmath}
\usepackage{amsfonts}
\usepackage{multirow}
\usepackage{booktabs}
\usepackage{bm}       
\usepackage{dcolumn} 
\usepackage{titlesec}
\usepackage[round-mode=places,round-precision=2]{siunitx}
\usepackage{blkarray}
\usepackage{color}
\usepackage{float}
\usepackage{graphicx}
\usepackage{tikz, tikz-3dplot}
\usepackage{subcaption}
\usepackage{caption}
\usepackage{mwe}
\usepackage{soul}
\usepackage{todonotes}
\usepackage{xr}
\usepackage[integrals]{wasysym}
\usepackage{braket}
\usepackage{cleveref}
\usepackage[normalem]{ulem}
\usepackage{tabularx}
\usepackage{titlesec}

\usepackage{setspace}

\setkeys{acs}{articletitle = true}
\setkeys{acs}{chaptertitle = true}
\setkeys{acs}{abbreviations = false}
\setkeys{acs}{maxauthors = 200}

\titleclass{\subsubsubsection}{straight}[\subsubsection]

\newcounter{subsubsubsection}[subsubsection]
\renewcommand\thesubsubsubsection{
  \thesubsubsection.\arabic{subsubsubsection}}

\titleformat{\subsubsubsection}
  {\normalfont\normalsize\bfseries}
  {\thesubsubsubsection}
  {1em}
  {}

\titlespacing*{\subsubsubsection}
  {0pt}{3.25ex plus 1ex minus .2ex}{1.5ex plus .2ex}

\crefname{subsubsubsection}{Sec.}{Secs.}

\crefname{equation}{eqn.}{eqns.}      
\Crefname{equation}{Eqn.}{Eqns.}

\crefname{section}{Sec.}{Secs.}      

\crefname{table}{Tab.}{Tabs.}   

\crefname{figure}{Fig.}{Figs.}

\newcommand{\Hi}{\hat{H}_{\mathrm{I}}}
\newcommand{\He}{\hat{H}_{\mathrm{E}}}
\newcommand{\Hie}{\hat{H}_{\mathrm{IE}}}

\newcommand{\V}{\hat{V}}

\newcommand{\tr}{\mathrm{Tr}_{\mathrm{E}}}
\newcommand{\den}{\hat{\rho}}  
\newcommand{\denmc}{\den_{\mathrm{mc}}}

\newcommand{\E}{\mathcal{E}}

\newcommand{\cre}[1]{a_{#1}^{\dag}}
\newcommand{\ann}[1]{a_{#1}}

\newcommand{\EF}{\varepsilon_{\mathrm{F}}}
\newcommand{\kB}{k_{\mathrm{B}}}

\author{Ellen T. Ekstrøm}
\altaffiliation{Equal contributions}
\affiliation{Department of Chemistry, Norwegian University of Science and Technology, N-7491 Trondheim, Norway}

\author{Jacob Pedersen}
\altaffiliation{Equal contributions}
\affiliation{Department of Chemistry, Norwegian University of Science and Technology, N-7491 Trondheim, Norway}

\author{Ida-Marie Høyvik}
\affiliation{Department of Chemistry, Norwegian University of Science and Technology, N-7491 Trondheim, Norway}
\email{ida-marie.hoyvik@ntnu.no}

\title{\Large Different Environments in Quantum Statistical Mechanics: \\ To $\beta$ or not to $(\kB T)^{-1}$}

\begin{document}

\maketitle

\begin{abstract}
    \noindent 
    We piece together textbook material to re-derive the parameter $\beta$ entering reduced density operators from quantum statistical mechanics. By re-deriving $\beta$, we show that the content of $\beta$ depends on the particular application of statistical mechanics. We focus on two different applications, namely, electronic-structure theory (new) and thermodynamics (standard). Specifically, we show that $\beta$ becomes proportional to the inverse Fermi energy, when the electronic states of a system interact with the infinitely many valence electrons in a metal. On the other hand, when the environment is a heat bath, $\beta$ takes the well-known form of inverse temperature. To highlight the importance of using the correct form of $\beta$, we explore statistical descriptions of a potassium atom adsorbed on a gold surface, where the valence electrons of gold represent the environment of potassium. By treating $\beta$ as the inverse Fermi energy, we are able to qualitatively reproduce the fractional charging of potassium. In contrast, if we interpret $\beta$ as the inverse temperature, extremely high and unphysical temperatures are required to reproduce the same qualitative picture. Thereby, we illustrate the fallacy of  not separating the mathematical framework of quantum statistical mechanics from its pervasive thermodynamic application.
\end{abstract}

\newpage
\section{Introduction}

The term ``statistical mechanics'' was coined by Gibbs,\cite{gibbs1902} but the field has its origin prior to that, in the kinetic theory of gases. Here, 
statistics and probabilities were used to handle an otherwise insurmountable number of degrees of freedom.\cite{ter1955foundations} Gibbs used the concept of an ensemble (a large number of identical, non-interacting copies of a system) ``\textit{to give precision to notions of probability}.''\cite{gibbs1902} Many attempts on the foundational ideas and the justification of statistical mechanics have been presented (see, e.g., the detailed review in Ref.~\citenum{ter1955foundations}). Among them, especially Tolman,\cite{tolman1979principles,tolman1940establishment} ter Haar,\cite{ter1955foundations} and Jaynes\cite{jaynes1957informationI,jaynes1957informationII,jaynes1965entropies} advocated for using statistical mechanics as a mathematical tool to make predictions about a system of interest based on the available knowledge. In other words, the ensembles become representative of different physical situations, e.g., isolated systems (the microcanonical ensemble), closed systems at equilibrium (the canonical ensemble), or open systems at equilibrium (the grand canonical ensemble).

Despite being developed prior to quantum mechanics, statistical mechanics may be used to describe quantum systems, e.g., at thermal equilibrium. In quantum statistical mechanics, the derivation of the ensembles start from the Schrödinger\cite{schrodinger1926undulatory} and the Liouville--von Neumann\cite{vonNeumann1932,von2018mathematical} equations. Here, the ensembles (through the corresponding density operators) encode the likelihood of the system being in different quantum states, and these are typically called ensemble weights or probabilities. Statistical mechanics is a framework able to determine the ensemble weights, and when the objective is a (quantum) state at thermal equilibrium, it is necessary to invoke Boltzmann's definition of entropy\cite{boltzmann1877beziehung,sharp2015translation,planck1978gesetz,planck1914heat} and the thermodynamic definition of temperature.\cite{callen1985thermodynamics,kittel1980thermal} However, this can be considered a specific \emph{application} of the \emph{mathematical framework} of statistical mechanics to thermodynamics.\cite{tolman1922relation,Khinchin1949mathematical}

The standard derivations of the canonical ensemble (and density operator) in quantum statistical mechanics can be found in many textbooks, e.g., Refs.~\citenum{zubarev,feynman1972statmech,balescu1975equilibrium,hill1987statmech,valkunas2013molecular}. The canonical ensemble can be derived from a microcanonical consideration of a subsystem of interest (I) and everything else, called the environment (E). The environment is usually assumed to be a heat bath, i.e., the environment Hamiltonian is considered a free particle model representing a monoatomic ideal gas. By averaging out the environment, one obtains the canonical density operator $\den_{\mathrm{c}}=Z^{-1}\exp(-\beta\Hi) $, where $Z$ is the partition function, and $\Hi$ is the Hamiltonian of the subsystem of interest. $\beta$ is defined as the energy derivative of the natural logarithm of the number of quantum states ($\mathrm{d} \ln{\Omega}_\mathrm{E} /\mathrm{d} E$) and contains all information about the environment. At this point, for a heat bath, one may use Boltzmann's definition of entropy ($S = \kB\ln\Omega$, where $k_{\mathrm{B}}$ is Boltzmann's constant),\cite{boltzmann1877beziehung,sharp2015translation,planck1978gesetz,planck1914heat} and the thermodynamic definition of temperature ($\mathrm{d}S/\mathrm{d}E= 1/T$).\cite{callen1985thermodynamics,kittel1980thermal} This yields a thermally equilibrated state characterized by $\beta =(k_{\mathrm{B}}T)^{-1}$. An alternative is to derive an expression for the number of quantum states (based on the free particle model) and calculate $(\mathrm{d} \ln{\Omega}_{\mathrm{E}}/\mathrm{d}E)$ analytically, arriving at an expression for $\beta$ which is proportional to the inverse average energy of the environment particles.\cite{feynman1972statmech} The average energy may be related to temperature through the equipartition theorem, whereby we once again arrive at the expression $\beta=(k_{\mathrm{B}}T)^{-1}$.

However, our interest lies not in thermalized states. We are interested in a quantum statistical treatment of how the electronic ground state of a quantum system is altered when it interacts with infinitely many (valence) electrons in the environment, as is the case when an atom or a molecule is adsorbed on a metal surface. This is an equilibrium situation, since we are seeking a stationary solution to the electronic Liouville--von Neumann equation for the adsorbed subsystem. The valence electrons in the metal can be described by a free particle model (i.e., as a free electron gas), and we may therefore exactly follow the standard derivation for thermalized states, which also utilize a free particle Hamiltonian. However, electronic-structure theory is temperature-free, and there is no need to invoke the concept of entropy, nor the thermodynamic definition of temperature.

None of the derivations presented in this work are new. The purpose of re-iterating what has been known for a long time,\cite{tolman1979principles} is that the application of statistical mechanics to thermodynamics is so common that it appears controversial to propose that $\beta$ depends on the physical setting (specifically, the environment Hamiltonian). In most chemistry textbooks, for example, the framework of statistical mechanics is intimately linked to the thermodynamic world, where the relationship between energy, work, temperature, and heat is the objective.\cite{uffink2007compendium,hill1986thermo,atkins2014quanta} This is quite natural as thermodynamics is a field of great importance in chemistry, making it a quintessential application of statistical mechanics. However, when describing the statistics of infinitely many electronic degrees of freedom interacting with the electronic states of an atom or a molecule, the environment Hamiltonian, and thus $\beta$, must reflect this.

In this work, we therefore provide a pedagogical exposition of how quantum statistical mechanics can have applications in equilibrium electronic-structure theory without any reference to thermal equilibrium. We show that using an environment Hamiltonian describing valence electrons in a metal (free electron gas) rather than a heat bath (monoatomic ideal gas), $\beta$ becomes proportional to the inverse Fermi energy. With this form of $\beta$ we can qualitatively reproduce the correct physical effects of the electronic states of potassium atoms adsorbed on a gold surface, in which the potassium atoms are seen experimentally\cite{farkas2009activation} and computationally\cite{ren2019k} to be fractionally charged. Using rather the wrong environment Hamiltonian (heat bath), temperature must be turned up to completely unphysical values to reproduce the same qualitative picture (illustrated in \cref{sec:wrong}). This is well-known, since for physical temperatures the energy-spacing of the electronic states is so large, that the electronic partition function is given by the degeneracy of the ground state.\cite{atkins2014quanta} We therefore think it is an important conceptual point to remember that $\beta$ need not be the inverse temperature, but rather, its meaning depends on the relevant environment. Of course, one may argue that the Fermi energy can also be converted to a ``temperature'' by a unit conversion using Boltzmann's constant. However, such conversion would do nothing besides obscuring the physical content of $\beta$ (p. 140 in Ref. \citenum{kittel2005ssp}).

The paper is organized as follows. In \cref{sec:box}, we present the theoretical framework and piece together standard derivations of the canonical ensemble and derive expressions of $\beta$ for two different types of environments. Next, we present an illustrative example, demonstrating that the grand canonical density operator is capable of reproducing the fractional charging of potassium atoms adsorbed on a gold surface when using the correct environment Hamiltonian (free electron gas) in \cref{sec:example}. Lastly, we summarize our work in \cref{sec:summary}.

\section{Theoretical Framework}\label{sec:box}

In this section, we piece together standard derivations of the density operators from quantum statistical mechanics necessary for discussing the meaning of $\beta$. We begin by motivating statistical descriptions and the use of density operators in \cref{sec:densityop}. We then consider isolated systems in \cref{sec:iso}. However, we are interested in atoms and molecules interacting with fermionic degrees of freedom, allowing the system of interest to be open to energy and possibly electron fluctuations. It quickly becomes prohibitively expensive to treat all degrees of freedom fully quantum mechanically when considering large environments. As an alternative, we may start from an  isolated description of the composite system (i.e., an atom or a molecule and its environment), and average out the environment. This leads to reduced descriptions as discussed in \cref{sec:red}. Here, $\beta$ enters as a parameter containing information about the environment. We provide a pedagogical exposition highlighting that the meaning of $\beta$ depends on the type of interacting environment. Having established the meaning of $\beta$, we review the standard reduced density operators in \cref{sec:qstatmech}.

\subsection{Statistical Descriptions}\label{sec:densityop}

To describe statistical uncertainty, we first review the concept of ensembles. An ensemble is a collection of a large number of identical, non-interacting copies of a system. From an electronic-structure theory perspective, each system may conveniently be specified by its (parametrically fixed) nuclear geometry, number of electrons, and electronic state. Here, copies in the ensemble refer to the same nuclear geometry, but they can differ in number of electrons and be in different electronic states. Each copy of the system is described by the electronic Schrödinger equation,\cite{schrodinger1926undulatory}
\begin{equation} \label{eq:SE}
    \hat{H} \ket{\Psi_{k}} = E_{k} \ket{\Psi_{k}}\,, \qquad k = 0,1,\dots\,,
\end{equation}
where $\hat{H}$ is the electronic Hamiltonian, and $\ket{\Psi_{k}}$ is the $k$th eigenstate with corresponding eigenenergy $E_{k}$. The electronic Hamiltonian (omitting nuclear repulsion) in the spin orbital basis is given by
\begin{equation}\label{eq:full_hamiltonian}
    \hat{H} = \sum_{PQ} h_{PQ} ~\cre{P} \ann{Q} + \frac{1}{2} \sum_{PQRS} g_{PQRS}~ \cre{P} \cre{R} \ann{S} \ann{Q} \,,
\end{equation} 
where $\cre{P} \ann{Q}$ and $\cre{P} \cre{R} \ann{S} \ann{Q}$ are one- and two-electron excitation operators, while $h_{PQ}$ and $g_{PQRS}$ are one- and two-electron integrals (detailed in, e.g., eqns.~(2.2.19) and (2.2.20) in Ref.~\citenum{MEST}), respectively. The number operator is given by
\begin{equation}
    \hat{N} = \sum_{P} \cre{P} \ann{P} \,.
\end{equation}

By means of the ensemble, we may define the statistical density operator as
\begin{equation} \label{eq:statden}
    \den = \sum_{k} w_{k} \ket{\Psi_{k}}\bra{\Psi_{k}} \,,
\end{equation}
where the summation runs over all eigenstates in the system, and the probabilities $\{w_{k}\}$ describe the likelihood of the system being in different electronic eigenstates. The probabilities are typically called ensemble weights and fulfill 
\begin{equation}\label{eq:physmeaningfull}
    \sum_{k} w_{k} = 1 \; , \qquad w_{k} \geq 0 \,.
\end{equation} 
In addition, the density operator (\cref{eq:statden}) must be Hermitian ($\hat{\rho} = \hat{\rho}^{\dagger}$) and, in equilibrium theory, a stationary solution to the Liouville--von Neumann equation, which means that
\begin{equation}\label{eq:lvn}
    [\hat{H}, \hat{\rho}] = 0 \,.
\end{equation}
The probabilities $\{w_{k}\}$ remain to be determined, which is the objective of statistical mechanics. In the following, we will establish relations for the ensemble weights in different situations.

\subsection{Isolated Systems}\label{sec:iso}

We start by considering an ensemble of isolated systems. By ``isolated'' we mean that the spread in both the energy and the number of electrons in the ensemble is zero,
\begin{align}
    \mathrm{Tr}\left\{\den\hat{H}^2 \right\} - \left(\mathrm{Tr}\left\{\den\hat{H}\right\}\right)^2 &=0 \label{eq:energyspreadmc} \\ 
    \mathrm{Tr}\left\{\den \hat{N}^2 \right\} - \left(\mathrm{Tr}\left\{\den\hat{N}\right\}\right)^2 &=0 \,.
\end{align}
We may insert the definition of the density operator (\cref{eq:statden}) in the energy-spread constraint (\cref{eq:energyspreadmc}) and evaluate the trace. The stationary condition in \cref{eq:lvn} means that $\den$ and $\hat{H}$ have simultaneous eigenstates, and by working in this eigenbasis, we obtain
\begin{equation}
    \sum_{k} w_{k} E_{k} \left( E_{k} - \sum_{l} w_{l} E_{l} \right) = 0 \,,
\end{equation}
which means that any non-trivial solution must fulfill
\begin{equation} 
     \sum_{l} w_{l} E_{l} = E_{k} \,.
\end{equation}
In other words, an average of the eigenenergies of the Hamiltonian must equal a single eigenenergy. By considering an isolated system in its ground state ($k=0$), we have
\begin{equation}\label{eq:equalapriori}
    \sum_{l} w_{l} E_{l} = E_{0} \,.
\end{equation}
Hence, to fulfill \cref{eq:equalapriori}, only states with energy $E_{0}$ may contribute to the mixture, as there by definition are no energetically lower-lying states than the ground state that may compensate for the inclusion of higher-lying states. In other words, only degenerate states are allowed to contribute
\begin{equation}
    \hat{H}\ket{\Psi_{k}}=E_{0}\ket{\Psi_{k}},  \qquad k = 0,1,\dots,\Omega(E_0)-1 \,,
\end{equation}
where $\Omega (E_0)$ is the number of degenerate states with energy $E_0$. Consequently, 
\begin{equation} \label{eq:mcweights}
    w_{k} = \Omega(E_{0})^{-1}, \qquad k = 0,1,\dots,\Omega(E_0)-1\,.
\end{equation} 
This is equivalent to the principle of equal \textit{a priori} probabilities, stating that the least biased assignment in the absence of any information to guide the distribution, is obtained by assigning equal probabilities to all involved states. We note that the equal \textit{a priori} probability distribution is the least biased way to fulfill the energy constraint (\cref{eq:energyspreadmc}) when considering a system in its ground state, rather than being a fundamental assumption or hypothesis. It is customary to distinguish between non-degenerate eigenstates $\Omega(E_0)=1$ and degenerate eigenstates $\Omega(E_0)>1$.

\subsubsection{Non-Degenerate Eigenstate} \label{sec:nondeg}

First, we consider the case of a non-degenerate ground state,
\begin{equation}
    \hat{H} \ket{\Psi_{0}} = E_{0} \ket{\Psi_{0}}\,.
\end{equation}
The probability of being in this state is 1. That is, the wave function contains the maximum-possible information about the system. The density operator for this state reads 
\begin{equation} \label{eq:pureden}
    \hat{\rho}_{\mathrm{pure}} = \ket{\Psi_{0}} \bra{\Psi_{0}} \,.
\end{equation}
We note that in addition to satisfying the trace and symmetry requirements discussed in \cref{sec:iso}, this operator is idempotent ($\hat{\rho}_{\mathrm{pure}}^{2} = \hat{\rho}_{\mathrm{pure}}$). Such a density operator is referred to as a pure state.

\subsubsection{Degenerate Eigenstates} \label{sec:deg}

Next, we consider the case of a multiply-degenerate ground state,
\begin{equation}
    \hat{H} \ket{\Psi_{k}} = E_{0} \ket{\Psi_{k}}\,,  \qquad k = 0, 1, ..., \Omega(E_{0}) -1 \,.
\end{equation}
The system will be a statistical mixture of the degenerate states, and the probabilities are given in \cref{eq:equalapriori}. By inserting \cref{eq:mcweights} in \cref{eq:statden} and using the resolution of identity ($1=\sum_{k} \ket{\Psi_{k}} \bra{\Psi_{k}}$), we obtain the microcanonical density operator,
\begin{equation} \label{eq:mcgeneral}
    \hat{\rho}_{\mathrm{mc}} = \Omega(E_{0})^{-1} \,.
\end{equation}
The microcanonical density operator fulfills the trace and symmetry requirements, but it is not idempotent. Rather, we have that $\hat{\rho}_{\mathrm{mc}}^2 \leq \hat{\rho}_{\mathrm{mc}}$, where the equality only holds for a non-degenerate state. The pure state density operator (\cref{eq:pureden}) may therefore be thought of as a special case of the microcanonical density operator.

\subsection{Reduced Descriptions} \label{sec:red}

The systems we are interested in (atoms and molecules) interact with an environment, with which it may share energy and/or electrons. The environment (E) represents everything else than the subsystem of interest (I). We may treat the combined subsystem of interest and environment as an isolated (composite) system. However, when the environment is large (e.g., infinitely many valence electrons in a metal), it is impossible to treat all degrees of freedom fully quantum mechanically. Rather than explicitly describing the composite system, we may focus on the subsystem of interest alone by averaging out the environmental degrees of freedom. The result is a reduced description, in which the degrees of freedom of the subsystem of interest are treated explicitly, while the environmental degrees of freedom are accounted for implicitly. However, we lose information about the subsystem of interest (specifically, details about the interaction) by averaging out the environment. To compensate for that, we must resort to a statistical treatment.

\subsubsection{Interacting Subsystems} \label{sec:partitioning}

The subsystem of interest and environment is assumed to have no shared orbitals (i.e., no covalent bonds). Thereby, we may spatially localize the spin orbitals of the composite system $\{ \varphi_{P} \}$ such that they may be categorized as belonging to either the subsystem of interest $\{ \varphi_{p} \}$ or the environment $\{ \varphi_{\bar p} \}$. In such local spin orbital basis ($\{ \varphi_{P} \} = \{ \varphi_{p} \} \cup \{ \varphi_{\bar p} \}$), the electronic Hamiltonian (\cref{eq:full_hamiltonian}) naturally partitions into 
\begin{equation}\label{eq:oqs}
    \hat{H} = \Hi + \He + \Hie \,, 
\end{equation}
where $\Hi$ and $\He$ describe the isolated subsystem of interest and the isolated environment, respectively. $\Hie$ describes the interaction between the subsystem of interest and the environment. 

We assume to construct eigenstates for the isolated subsystem of interest and the isolated environment,
\begin{align}
    \Hi \ket{\Psi_{n}} &= E_{n} \ket{\Psi_{n}}\,, \qquad n = 0,1,\dots\label{eq:syseig} \\
    \He \ket{\Psi_{\bar n}} &= E_{\bar n} \ket{\Psi_{\bar n}}\,, \qquad \Bar{n} = 0,1,\dots \label{eq:enveig}\,,
\end{align}
where unbarred and barred indices distinguish between the subsystem of interest and the environment, respectively. Similarly, the number operator of the composite system splits into a subsystem of interest and an environment number operator, $\hat{N} = \hat{N}_{\mathrm{I}} + \hat{N}_{\mathrm{E}}$, such that
\begin{align}
    \hat{N}_{\mathrm{I}}\ket{\Psi_{n}} &= N_{\mathrm{I}}\ket{\Psi_{n}} \label{eq:sysnum}\\
    \hat{N}_{\mathrm{E}}\ket{\Psi_{\Bar{n}}} &= N_{\mathrm{E}}\ket{\Psi_{\Bar{n}}} \label{eq:envnum} \,.
\end{align}
Additional details and discussions on our partitioning of the composite system may be found in Refs.\citenum{sannes2025fractional} and \citenum{pedersen2025quantum}.

\subsubsection{Energy Fluctuations}\label{sec:fluc}

The interaction between the subsystem of interest and the environment leads to energy and/or electron-number fluctuations. In this section, we re-derive a reduced density operator able to model such energy fluctuations, while conserving the number of electrons within each of the two subsystems. We closely follow the derivation as presented by Zubarev.\cite{zubarev} The energy fluctuations are caused by the interaction Hamiltonian $\Hie$ in \cref{eq:oqs}. Nonetheless, it is standard practice to neglect the interaction Hamiltonian under the assumption that the interaction is weak,
\begin{equation}\label{eq:canonical-hamil}
    \hat{H} \approx \Hi + \He\,.
\end{equation}
This new composite Hamiltonian is additive separable and the corresponding composite states are multiplicative separable. The subsystem of interest and environment Hamiltonians commute, and we may therefore construct simultaneous eigenstates from \cref{eq:syseig,eq:enveig},
\begin{equation}\label{eq:compositestate}
    (\Hi + \He) \ket{\Psi_{n} \Psi_{\bar n}} = (E_{n} + E_{\bar n}) \ket{\Psi_{n} \Psi_{\bar n}} \,.
\end{equation}
Neither the subsystem of interest nor the environment Hamiltonian commute with the interaction Hamiltonian, and the product states in \cref{eq:compositestate} are thus not eigenstates of the composite electronic Hamiltonian (\cref{eq:oqs}). We denote the approximate (ground state) energy of the composite system as
\begin{equation}\label{eq:tot-energy}
    \E \approx E_{n} + E_{\bar n} \,.
\end{equation}
To compensate for neglecting the interaction Hamiltonian (responsible for the phenomenon that we aim to model), it is customary to allow a small non-zero spread in the energy of the two subsystems, as this would have been the effect of the interaction Hamiltonian. State degeneracies in the composite system may now originate from coupling of the available states in the subsystem of interest with environmental states under the condition that \cref{eq:tot-energy} is satisfied.

The density operator for the composite system takes the form of the microcanonical density operator from \cref{eq:mcgeneral},
\begin{equation}
    \denmc=\sum_{n\Bar{n}} w_{\mathrm{mc}}(E_{n} + E_{\bar n})
    \ket{\Psi_{n} \Psi_{\bar n}}\bra{\Psi_{n} \Psi_{\bar n}}\,,
\end{equation}
where the summation runs over $n$ and $\bar n$ such that $ E_{n} + E_{\bar n}\approx \E$. The coupling of states is a combinatorial problem, and we may write the number of states in the composite system (i.e., the degeneracy) as 
\begin{equation}\label{eq:combi}
    \Omega (\E) = \sum_{n}\Omega_{\mathrm{I}}(E_{n})\Omega_{\mathrm{E}}(\E - E_{n}) \,,
\end{equation}
where $\Omega_{\mathrm{I}}(E_{n})$ is the number of states in the subsystem of interest with energy $E_{n}$, and $\Omega_{\mathrm{E}}(\E - E_{n})$ is the number of states in the environment with energy $\E - E_{n} = E_{\bar n}$ (\cref{eq:tot-energy}).

A reduced description of the system is obtained by averaging out the environment,
\begin{equation}\label{eq:canonical-density-operator}
    \den_{\mathrm{c}} = \tr\,\{\denmc\} = \sum_{n} w_{\mathrm{c}}(E_n) \ket{\Psi_{n}}\bra{\Psi_{n}} \,,
\end{equation} 
with the weights,
\begin{equation} \label{eq:canstartweight}
    w_{\mathrm{c}}(E_n) 
    = \sum_{\Bar{n}} w_{\mathrm{mc}}(E_n + E_{\Bar{n}})  = \, \frac{1}{\Omega(\E)} \sum_{\Bar{n}} \; 1 = \, \frac{\Omega_{\mathrm{E}}(\E-E_n)}{\Omega(\E)} \,,
\end{equation}
where we have used \cref{eq:mcgeneral} to write the second equality. The last summation counts the number of states in the environment. $\tr\{\cdot\}$ denotes the partial trace over the environmental degrees of freedom. The partial trace requires special attention when considering an electronic environment to avoid the fermionic partial trace ambiguity (i.e., electrons losing their electronic behavior).\cite{montero2011fermionic-ambiguity, friis2013fermionic} We have recently presented a definition of the fermionic partial trace operation that resolves this ambiguity. More details are given in Ref.~\citenum{pedersen2025quantum}. The important point for the present work is that the fermionic partial trace is well-defined.

The reduced density operator in \cref{eq:canonical-density-operator} is called the canonical density operator, and it describes a subsystem of interest open to energy fluctuations with its environment. We note that such a subsystem is typically referred to as ``closed''. We may mathematically define ``closed'' as having a non-zero energy-spread, while retaining a zero spread in the electron-number,
\begin{align}
    \mathrm{Tr}\left\{\den_{\mathrm{c}}\Hi^2 \right\} - \left(\mathrm{Tr}\left\{\den_{\mathrm{c}}\Hi\right\}\right)^2 &\neq 0  \label{eq:canspread} \\ 
    \mathrm{Tr}\left\{\den_{\mathrm{c}} \hat{N}_{\mathrm{I}}^2 \right\} - \left(\mathrm{Tr}\left\{\den_{\mathrm{c}}\hat{N}_{\mathrm{I}}\right\}\right)^2 &=0 \,.
\end{align}
Furthermore, we note that the subsystem of interest remain at statistical equilibrium with the constant average energy,
\begin{equation} \label{eq:canenergy}
    \langle E_{\mathrm{I}}\rangle = \mathrm{Tr}\left\{\den_{\mathrm{c}}\Hi \right\}\,,
\end{equation}
but $\langle E_{\mathrm{I}}\rangle$ is not an eigenvalue of $\Hi$. 

The canonical ensemble weights in \cref{eq:canstartweight} may be written as
\begin{equation}\label{eq:c-weights-NoE}
    \begin{split}
        w_{\mathrm{c}}(E_n)=& 
        \;\frac{\Omega_{\mathrm{E}}(\E-E_n)}{\sum_{m}\Omega_{\mathrm{I}}(E_m)\Omega_{\mathrm{E}}(\E-E_m)}
        \approx  \;\frac{\exp(\ln{\Omega_{\mathrm{E}}(\E-E_n)})}{\sum_{m}\exp(\ln{\Omega_{\mathrm{E}}(\E-E_m)})} \,,
    \end{split}
\end{equation}
where we have used \cref{eq:combi} to write the first equality. The last equality is obtained by (i) using the logarithmic product rule, $\ln\Omega_{\mathrm{I}}(E_m)\Omega_{\mathrm{E}}(\E-E_m) = \ln\Omega_{\mathrm{I}}(E_m) + \ln\Omega_{\mathrm{E}}(\E-E_m)$, and (ii) assuming that the environment is significantly larger than the subsystem of interest ($\Omega_{\mathrm{I}}(E_m) \ll \Omega_{\mathrm{E}}(\E-E_m)$). Thereby, the contribution from the subsystem of interest can be neglected.

The exponential can now be expanded in a Taylor series about $\E$, and by retaining terms up to first order in $E_n$, we obtain
\begin{equation}\label{eq:ln-taylor}
    \ln{\Omega_{\mathrm{E}}(\E-E_n)} \simeq \ln{\Omega_{\mathrm{E}}(\E)}  - \frac{\mathrm{d}\ln\Omega_{\mathrm{E}}\left(\E \right)}{\mathrm{d}\E} E_n \, ,
\end{equation}
where the energy derivative is assumed constant in the energy range under consideration (p. 3 in Ref. \citenum{feynman1972statmech}). $\beta$ may now be defined as the energy derivative,\cite{feynman1972statmech,zubarev,balescu1975equilibrium}
\begin{equation}\label{eq:beta-zubarev}
    \beta = \frac{\mathrm{d}\ln\Omega_{\mathrm{E}}\left(\E \right)}{\mathrm{d}\E} \,.
\end{equation}
The canonical density operator may be obtained by combining \cref{eq:c-weights-NoE,eq:ln-taylor,eq:beta-zubarev} with \cref{eq:canonical-density-operator}, as will be discussed in \cref{sec:qstatmech}. Before that, we need to derive explicit expressions for $\beta$, and this requires an analytical expression for the number of states in the environment. The free particle model is commonly used for that, and this model allows for the treatment of both valence electrons in a metal (i.e., free electron gas) and a heat bath (i.e., monoatomic ideal gas).

\subsubsection{Free Particle Model}\label{sec:free}

Feynman\cite{feynman1972statmech} relates $\beta$ to the inverse average energy of the environment by modeling the environment as a collection of independent, non-interacting free particles in a box. In this section, we briefly review the free particle model and show applications to two different free particle environments.  

A particle in the environment may be described by the time-independent, one-particle Schrödinger equation,
\begin{equation}\label{eq:schr-pib}
    \He^{\mathrm{1p}}\ket{\varphi_{\Bar{n}}}=\varepsilon_{\Bar{n}}\ket{\varphi_{\Bar{n}}}\,, \qquad \Bar{n} = 1,2,\dots \,, 
\end{equation}
where $\He^{\mathrm{1p}}$ is the one-particle Hamiltonian (including units) given by
\begin{equation} \label{eq:singleH}
    \He^{\mathrm{1p}} = -\frac{\hbar^2}{2m}\hat{\nabla}^2 + \V \,,
\end{equation}
with $m$ being the mass of the particle. We note that $\ket{\varphi_{\Bar{n}}}$ in \cref{eq:schr-pib}  is a spin orbital, and that $\Bar{n}$ runs from 1, as it reflects the quantum numbers for the free particle model. The particle is assumed to be confined within a volume $\mathcal{V}_{\mathrm{box}}$, wherein it can move freely. In other words, $ \V = 0$ inside the box, and $ \V = \infty$ everywhere else. The corresponding eigenenergy is 
\begin{equation}\label{eq:e-3d-pib}
    \varepsilon_{\Bar{n}} = \frac{\hbar^2 }{2m}\,k_{\Bar{n}}^{2}\,,
\end{equation}
where $k_{\Bar{n}}$ is the wave vector of the wave function. It may be shown (e.g., eqn.~(20) in Ref.~\citenum{kittel2005ssp}, eqn.~(2.61) in Ref.~\citenum{ashcroft1976solid}, eqn.~(7.51) in Ref. \citenum{atkins2011molecular}, or eqn.~(3.164) in Ref.~\citenum{patterson2010ssp}) that the one-particle density of states is given by
\begin{equation}\label{eq:dos}
    \eta (\varepsilon) = \frac{\mathcal{V}_{\mathrm{box}}}{4\pi^2}\left(\frac{2 m}{\hbar^2}\right)^{3/2}\sqrt{\varepsilon}\,.
\end{equation}
The number of states for a single particle in the environment may be computed as
\begin{equation}
    \Omega_{\mathrm{E}}^{\mathrm{1p}}(\E) = \int_{0}^{\E} \eta(\varepsilon) \, \mathrm{d}\varepsilon 
    = \frac{\mathcal{V}_{\mathrm{box}}}{4\pi^2}\left(\frac{2 m}{\hbar^2}\right)^{3/2} (\sqrt{\E})^{3}\,,
\end{equation}
where we have used that the environment is much larger than the subsystem of interest (compatible with the approximations invoked in \cref{sec:fluc}), such that the energy of the environment may be approximated as the ground state energy of the composite system ($E_{\Bar{n}} \approx \E$). For $N_{\mathrm{E}}$ particles, the number of states  in the environment becomes
\begin{equation}\label{eq:dos-3d}
    \Omega_{\mathrm{E}}(\E) = \left(\frac{\mathcal{V}_{\mathrm{box}}}{4\pi^2}\left(\frac{2 m}{\hbar^2}\right)^{3/2}\right)^{N_{\mathrm{E}}}(\sqrt{\E})^{3N_{\mathrm{E}}}\,.
\end{equation}
We may now obtain $\beta$ by computing the energy derivative in \cref{eq:beta-zubarev} analytically, yielding
\begin{equation}\label{eq:beta}
    \beta = \frac{\mathrm{d}\ln\Omega_{\mathrm{E}}\left(\E \right)}{\mathrm{d}\E}= \frac{\mathrm{d}}{\mathrm{d}\E}\ln\,(\sqrt{\E})^{3N_{\mathrm{E}}}  = \frac{1}{\frac{2}{3}W} \,,
\end{equation}
where $W = \E/N_{\mathrm{E}}$ is defined as energy per particle in the last equality. This result relates $\beta$ to the inverse average energy, and it is irrespective of the type of environment (as long as it can be described by a free particle model). We will now consider two different environments, namely, a heat bath and an electronic environment.

\subsubsubsection{\small Statistical Mechanics for Thermodynamics: Monoatomic Ideal Gas} \label{sec:equi}

We first consider a heat bath, and we may obtain an expression for the energy per particle starting from Boltzmann's definition of entropy, 
\begin{equation}\label{eq:boltz}
    S = \kB \ln\Omega_{\mathrm{E}}(\E)\,.
\end{equation}
Specifically, by (i) inserting the number of states as obtained from the free particle model in \cref{eq:dos-3d}, (ii) evaluating the energy derivative, 
\begin{equation}
    \frac{\mathrm{d} S}{\mathrm{d} \E} =  \frac{3 \kB N_{\mathrm{E}}}{2\E}\,,
\end{equation}
and (iii) invoking the thermodynamic definition of temperature,
\begin{equation}\label{eq:tempe}
    \frac{\mathrm{d} S}{\mathrm{d} \E} = \frac{1}{T}\,,
\end{equation}
we obtain the equipartition theorem,\cite{atkins2014quanta}
\begin{equation}\label{eq:equipartition}
    W = \frac{\E}{N_{\mathrm{E}}} = \frac{3}{2} \kB T\,.
\end{equation}
By means of the equipartition theorem, $\beta$ in \cref{eq:beta} becomes\cite{feynman1972statmech,atkins2014quanta,balescu1975equilibrium,callen1985thermodynamics,hill1986thermo} 
\begin{equation}\label{eq:inversetemp}
    \beta =  (\kB T)^{-1} \,.
\end{equation}
In this specific application of statistical mechanics to thermodynamics, $\beta$ may be identified as the inverse temperature. We note that the same expression could have been obtained by directly inserting \cref{eq:boltz,eq:tempe} into \cref{eq:beta-zubarev}, but we show the derivation through the free particle model to highlight the role of the environment.

\subsubsubsection{\small Statistical Mechanics for Electronic-Structure Theory: Free Electron Gas}\label{sec:bound}

We will now consider the valence electrons in a metal as the environment. The valence electrons in the metal may be described by the free particle model in \cref{sec:free} (i.e., as a free electron gas), and we may therefore follow Feynman's approach and compute the energy per particle using an electronic (one-particle) environment Hamiltonian.

The ground state energy of the environment may be computed as the sum of all one-electron energies up to the topmost-filled level, which energy is referred to as the Fermi energy and is given by
\begin{equation}
    \EF = \frac{\hbar^2}{2m_{\mathrm{e}}}\,k_{\mathrm{F}}^{2} \,,
\end{equation}
where $m_{\mathrm{e}}$ is the mass of an electron, and $k_{\mathrm{F}}$ is the Fermi wave vector. For an infinitely large environment, the summation converges to an integral weighted by the density of states (\cref{eq:dos}). Hence, the total energy of the valence band becomes\cite{patterson2010ssp,ashcroft1976solid,atkins2011molecular}
\begin{equation}\label{eq:au-E-tot}
    \E = \int_{0}^{\EF} E\, \eta(E)\;\mathrm{d}E = \frac{\mathcal{V}_{\mathrm{box}}}{4\pi^2}\left(\frac{2 m_{\mathrm{e}}}{\hbar^2}\right)^{3/2}\frac{2}{5} \EF^{5/2} \,.
\end{equation}
Similarly, the number of electrons in the metal may be computed by integrating the one-particle density of states over the occupied spin orbitals (as there is one electron per spin orbital), resulting in 
\begin{equation}\label{eq:au-N-tot}
    N_{\mathrm{E}} = \int_{0}^{\EF} \eta(E) \; \mathrm{d}E = \frac{\mathcal{V}_{\mathrm{box}}}{4\pi^2}\left(\frac{2 m_{\mathrm{e}}}{\hbar^2}\right)^{3/2}\frac{2}{3} \EF^{3/2} \,.
\end{equation}
The energy per electron in a metal becomes\cite{widom2002statmech,ashcroft1976solid}
\begin{equation}
    W = \frac{3}{5}\EF \,.
\end{equation}
Consequently, $\beta$ in \cref{eq:beta-zubarev} becomes proportional to the inverse Fermi energy, 
\begin{equation}\label{eq:beta-metal}
    \beta = \frac{5}{2}\EF^{-1}\,.
\end{equation}

\subsubsubsection{\small Electronic Environment versus Heat Bath}
The choice of environment Hamiltonian depends on the type of environment. Both a heat bath and an electronic environment can be described with a free particle model using the one-particle Hamiltonian in \cref{eq:singleH}, in which $\beta$ may be interpreted as the inverse energy per particle (\cref{eq:beta}). The different meanings of $\beta$ in \cref{eq:inversetemp,eq:beta-metal} arises from the different applications of quantum statistical mechanics, whether that being to thermodynamics or electronic-structure theory. For a heat bath (i.e., monoatomic ideal gas), $\beta$ is identified as the inverse temperature. In contrast, for an electronic environment (i.e., free electron gas), $\beta$ is identified as the inverse Fermi energy. The Fermi energy may be converted to a ``temperature''  by a unit conversion using Boltzmann's constant. However, using a temperature unit to reflect the energy scale would do nothing besides obscuring the physical content of $\beta$.\cite{kittel2005ssp} For the same reason, we do not speak of ground state and excitation temperatures. Moreover, a comparison of \cref{eq:beta-metal} and \cref{eq:inversetemp} shows that simply converting the Fermi energy to a ``temperature'' fails to reproduce the correct prefactor.

Both environments may be described by the same one-particle Hamiltonian in \cref{eq:singleH}, with the key difference being the mass of their particles. Specifically, a heat bath consists of a monoatomic ideal gas (with masses dominated by the nuclei), wheres an electronic environment consists of infinitely many electrons. The spacing between the energy levels is inversely proportional to the mass of the particle (\cref{eq:e-3d-pib}). Therefore, an environment of heavier particles will have much smaller spacing between the energy levels. In the same line of thought, as $\beta$ is proportional to the inverse energy (\cref{eq:beta-metal}), it becomes proportional to the mass of the particle, meaning that heat baths will be associated with much larger values of $\beta$ than electronic environments. Consequently, if we (incorrectly) model an electronic environment as a monoatomic ideal gas, extremely high and unphysical temperatures are required to compensate for the much heavier particles (illustrated in \cref{sec:wrong}), and thereby lower the value of $\beta$.

\subsection{Reduced Density Operators} \label{sec:qstatmech}

In this section, we pick up from where we left in \cref{sec:fluc} and finish the derivation of the canonical density operator. Specifically, we insert the Taylor expansion in \cref{eq:ln-taylor} into the canonical ensemble weights in \cref{eq:c-weights-NoE}, split the exponentials, and simplify the expression. The result is
\begin{equation}
    \den_{\mathrm{c}} = Z^{-1}\exp(-\beta\hat{H}_{\mathrm{I}}) \,,
\end{equation}
where we have used the resolution of identity (i.e., $1=\sum_{n} \ket{\Psi_{n}} \bra{\Psi_{n}}$), and that $\ket{\Psi_{n}}$ is an eigenfunction of $\hat{H}_{\mathrm{I}}$ to re-introduce the Hamiltonian of the subsystem of interest. Moreover, we have defined the canonical partition function,
\begin{equation}
    Z = \mathrm{Tr}\left\{\exp(-\beta\hat{H}_{\mathrm{I}}) \right\}.
\end{equation}

In an experimental setting, it can be difficult to fix the particle number. Therefore, it may be convenient to consider a system that is open to both energy and electron-number fluctuations,
\begin{align}
    \mathrm{Tr}\left\{\den_{\mathrm{gc}}\Hi^2 \right\} - \left(\mathrm{Tr}\left\{\den_{\mathrm{gc}}\Hi\right\}\right)^2 &\neq 0 \\ 
    \mathrm{Tr}\left\{\den_{\mathrm{gc}} \hat{N}_{\mathrm{I}}^2 \right\} - \left(\mathrm{Tr}\left\{\den_{\mathrm{gc}}\hat{N}_{\mathrm{I}}\right\}\right)^2 &\neq0 \,.
\end{align}
We may derive such a reduced density operator following the same approach as for the canonical density operator. The result is the grand canonical density operator, and it takes the form\cite{feynman1972statmech,zubarev,balescu1975equilibrium}
\begin{equation}\label{eq:gc}
    \den_{\mathrm{gc}} = \Xi^{-1}\exp(-\beta(\hat{H}_{\mathrm{I}}-\mu\hat{N}_{\mathrm{I}})) \,,
\end{equation}
where $\Xi$ is the grand canonical partition function given by
\begin{equation}
    \Xi = \mathrm{Tr}\left\{\exp(-\beta(\hat{H}_{\mathrm{I}}-\mu\hat{N}_{\mathrm{I}})) \right\} \,,
\end{equation}
with $\mu$ being the chemical potential, describing the energy cost of moving an electron between the subsystem of interest and the environment. Recently, we have presented a reduced density operator that provides new insights into the chemical potential.\cite{pedersen2025quantum} However, this is beyond the scope of the present study.

\section{Illustrative Example} \label{sec:example}

To demonstrate how statistical mechanics may have qualitative applications in equilibrium electronic-structure theory, we now turn to the adsorption of potassium atoms on an Au(111) surface, where the potassium atoms have been observed experimentally\cite{farkas2009activation} and computationally\cite{ren2019k} to be fractionally charged. Specifically, Ren et al. reported that potassium donates approximately 0.8 electrons to the surface (based on density functional theory calculations and Bader's quantum theory of atoms in molecules\cite{bader1985atoms,bader1975molecular}), leaving the potassium atom with a fractional charge on average.\cite{ren2019k}. The fractional charging of potassium may be described by considering a statistical mixture of the cationic (K$^{+}$) and neutral (K) atoms. The probabilities (ensemble weights) of the charge states are given by the grand canonical density operator (\cref{eq:gc}), and we may write
\begin{align}
    w_{\mathrm{K}^{+}} &= \,\Xi^{-1} \exp (-\beta [ E_{N-1} - \mu (N-1) ] ) \label{eq:prob1}\\
    w_{\mathrm{K}} &= \,\Xi^{-1} \exp (-\beta [ E_{N} - \mu N ] ) \label{eq:prob2} \,. 
\end{align}
We assume that no other charge states have a significant contribution to the adsorption, meaning that
\begin{equation}
    w_{\mathrm{K}^{+}} + w_{\mathrm{K}} = 1 \,.
\end{equation}
The probabilities may be computed using either purely experimentally obtained quantities or computed values as shown in \cref{sec:egas}. Here, we recall that the statistical treatment is not intended (nor expected) to outperform full electronic-structure calculations. Rather, we expect a rough estimate, able to capture the correct physics.

The ratio between the probabilities of the cationic ($w_{\mathrm{K}^+}$) and the neutral ($w_{\mathrm{K}}$) potassium atom can be written as
\begin{equation}
    \frac{w_{\mathrm{K}^{+}}}{w_{\mathrm{K}}} = \exp\left(\beta[-\mu -(E_{N-1}-E_{N})] \right) \,,
\end{equation}
where we recognize the ionization energy of the potassium atom, $I = E_{N-1} - E_{N}$. The chemical potential describes the energy cost of moving an electron between the subsystem of interest and the environment. The work function $\Phi_{\mathrm{WF}}$ amounts to the energy required to remove an electron from the surface, and as we are interested in describing the donation of an electron from the potassium atom to the gold surface, we may approximate the negative chemical potential as the work function, $-\mu=\Phi_{\mathrm{WF}}$. Hence, the ratio between the probabilities can be written in terms of the ionization energy of potassium and the work function of gold as
\begin{equation}\label{eq:ratio}
    \frac{w_{\mathrm{K}^{+}}}{w_{\mathrm{K}}} = \exp \left( \beta [\Phi_{\mathrm{WF}} - I] \right) \,.
\end{equation}

\subsection{Electronic Environment Hamiltonian}\label{sec:egas}

In this section, we use $\beta$ as defined in \cref{eq:beta-metal}, and within the single electron picture, we may approximate the Fermi energy as the work function, $\EF = \Phi_{\mathrm{WF}}$.\cite{koopmans1934zuordnung} Here, it is important to note that the energy of a bound electron is negative, but the Fermi energy is a shifted quantity (relative to the potential in \cref{eq:singleH}). In other words, the definition of the work function assumes that the energy of the level wherefrom we are removing an electron is negative, but this minus sign is already included in the Fermi energy (making it a positive quantity).

We may compute the ratio in \cref{eq:ratio} using experimental quantities. The work function for Au(111) is found to be 5.31 eV,\cite{michaelson1977work} and the ionization energy of potassium to be 4.34 eV.\cite{lorenzen1981precise} We obtain $w_{\mathrm{K}^{+}} = 0.61$ and $w_{\mathrm{K}} = 0.39$, which is qualitatively comparable to that reported by Ren et al.\cite{ren2019k}

Alternatively, we may compute the probabilities directly from \cref{eq:prob1,eq:prob2}. For that purpose, we take $-\mu=\EF$,\cite{schulte1974chemical} and the Fermi energy of gold is found to be 5.51 eV.\cite{kittel2005ssp} The energies of K and K$^+$ are computed at the unrestricted and restricted Hartree--Fock level of theory,\cite{hartree1928waveI,hartree1928waveII,slater1930note,fock1930naherungsmethode,slater1928self,gaunt1928theory,pople1954self} respectively, using the 6-31G basis set\cite{rassolov19986} in the $e^{T}$ electronic-structure program.\cite{folkestad20262} We get $w_{\mathrm{K}^{+}} = 0.66$ and $w_{\mathrm{K}} = 0.34$, which is similar to that found using experimental values.

\subsection{Heat Bath Hamiltonian} \label{sec:wrong}
Here, we use the thermodynamic version of $\beta$ in \cref{eq:inversetemp} to illustrate the fallacy of not separating the mathematical framework of quantum statistical mechanics from its pervasive thermodynamic application. In other words, we (wrongly) treat the valence electrons in the metal as a heat bath.

The probabilities are computed at various temperatures, and the results are compiled in \cref{tab:probs}. A pure state of the cationic potassium atom is predicted at room temperature, and we find that a temperature on the order of 2 500 K is required to yield a statistical mixture including just 1\% of the neutral state. To reach the same probability distribution as computed in \cref{sec:egas}, temperatures around 25 000 K are needed. However, there will neither be potassium atoms nor a gold surface at such elevated temperatures, but instead a hot plasma. This makes the notion of temperature awkward, and is (or at least, should be) a strong indicator that temperature \textit{per se} has no place in quantum statistical mechanics for electronic-structure applications.

\begin{table}[H]
    \centering
    \begin{tabular}{rcc}
        \hline
        \hline \\
        $T$ [K] & $w_{\mathrm{K}^{+}}$ & $w_{\mathrm{K}}$ \\
        \\
        \hline
        \\
        298 & 1.00 & 0.00 \\
        2 500 & 0.99 & 0.01 \\
        5 000 & 0.90 & 0.10 \\
        7 500 & 0.82 & 0.18 \\
        10 000 & 0.76 & 0.24 \\
        15 000 & 0.68 & 0.32 \\
        20 000 & 0.64 & 0.36 \\
        25 000 & 0.61 & 0.39 \\
        30 000 & 0.59 & 0.41 \\
        \\
         \hline
         \hline
    \end{tabular}
    \captionsetup{width=0.85\textwidth, font={stretch=1.5}}
    \caption{Computed probabilities of the cationic $w_{\mathrm{K}^{+}}$ and neutral $w_{\mathrm{K}}$ atom using \cref{eq:ratio} with $\beta$ as given in \cref{eq:inversetemp} at various temperatures $T$.}
    \label{tab:probs}
\end{table}

\newpage
\section{Summary} \label{sec:summary}

In this work, we piece together the textbook derivations of the canonical density operator as presented by Zubarev\cite{zubarev} and Feynman\cite{feynman1972statmech} to discuss the meaning of $\beta$. Specifically, by (i) starting from a microcanonical description of the combined subsystem of interest and the environment (while allowing for small energy fluctuations between them) and (ii) averaging out the environment, one obtains the canonical density operator. All remaining information about the environment is contained within $\beta$, defined as the energy derivative of the natural logarithm of the number of quantum states in the environment. Next, by (iii) deriving an expression for the number of quantum states (based on the free particle model) and (iv) calculating the energy derivative analytically, $\beta$ may be related to the inverse energy per particle in the environment. This is irrespective of the specifics of the environment, as long as the environment may be described by a free particle model.

However, the energy per particle depends on the type of environment. If we assume a heat bath (i.e., a monoatomic ideal gas), we may invoke the equipartition theorem, thereby relating the average kinetic energy of the particles to a temperature. We emphasize that the equipartition theorem relies on Boltzmann's definition of entropy as well as the thermodynamic definition of temperature. Hence, the standard definition of $\beta$ as the inverse temperature necessitates a thermodynamic setting (i.e., thermal equilibrium). If instead the environment is the valence electrons in a metal (i.e., a free electron gas), we have to compute the average energy using an electronic free particle Hamiltonian. We show that $\beta$ in that case becomes proportional to the inverse Fermi energy of the metal. Moreover, we note that it is always possible to convert an energy to a ``temperature'' by a unit conversion using Boltzmann's constant, but such conversion would do nothing besides obscuring the physical content of $\beta$.

By treating $\beta$ as the inverse Fermi energy, we qualitatively reproduce the fractional charging of potassium atoms adsorbed on a gold surface. If we instead use the wrong environment Hamiltonian (for the heat bath), extremely high and unphysical temperatures are required to overcome the large energy-spacing between the electronic states, and thereby reproduce the same qualitative picture. Ultimately, we highlight the key conceptual point of this work, namely that the content of $\beta$ depends on the application of quantum statistical mechanics, and more specifically, the type of environment.

\section{Acknowledgment}

E.T.E, J.P., and I.-M.H. acknowledge funding from the European Research Council (ERC, OpenQuantum, 101170817). We thank Per-Olof Åstrand and Tore Haug-Warberg for valuable discussions.

\bibliography{references}

\end{document}